# Stabilization of single- and multi-peak solitons in the fractional nonlinear Schrödinger equation with a trapping potential

Yunli Qiu[1], Boris A. Malomed[2,3], Dumitru Mihalache[4], Xing Zhu[1], Xi Peng[1], and Yingji He[1]*

[1] *School of Photoelectric Engineering, Guangdong Polytechnic Normal University, Guangzhou 510665, China*

[2] *Department of Physical Electronics, School of Electrical Engineering, Faculty of Engineering, and Center for Light-Matter Interaction, Tel Aviv University, Tel Aviv 69978, Israel*

[3] *Instituto de Alta Investigación, Universidad de Tarapacá, Casilla 7D, Arica, Chile*

[4] *HoriaHulubei National Institute for Physics and Nuclear Engineering, P.O. Box MG-6, RO-077125, Bucharest-Magurele, Romania*

*Corresponding author: heyingji8@126.com

We address the existence and stability of localized modes in the framework of the fractional nonlinear Schrödinger equation (FNSE) with the focusing cubic or focusing-defocusing cubic-quintic nonlinearity and a confining harmonic-oscillator (HO) potential. Approximate analytical solutions are obtained in the form of Hermite-Gauss modes. The linear stability analysis and direct simulations reveal that, under the action of the cubic self-focusing, the single-peak ground state and the dipole mode are stabilized by the HO potential at values of the Lévy index (the fractionality degree) $\alpha \leq 1$, which lead to the critical or supercritical collapse in free space. In addition to that, the inclusion of the quintic self-defocusing provides stabilization of higher-order modes, with the number of local peaks up to seven, at least.

## 1. INTRODUCTION

Realizations of fractional nonlinear Schrödinger equations (FNSEs), which were first proposed by Laskin [1], have drawn much interest in various areas of physics [2-7]. In particular, an implementation of FNSE in spherical optical cavities for the generation of dual Airy beams was elaborated by Longhi [8]. In terms of the linear Schrödinger equation with a harmonic-oscillator (HO) potential, propagation of chirped Gaussian beams was investigated analytically and numerically [9]. Then, several types of solitons trapped in the HO potential were reported, assuming that the respective Lévy index (LI), which characterizes the fractionality of the equation [see Eq. (1) below], belongs to interval $1 < \alpha \leq 2$ [10,11] (the usual cubic nonlinear Schrödinger equation corresponds to $\alpha = 2$). It was also found that LI can control the splitting of Airy beams in the linear free-space fractional Schrödinger equation (without a potential), and the effect of periodic self-imaging under the action of a symmetric potential barrier was predicted for $\alpha = 1$ [12]. Anomalous interactions of Airy waves in the framework of FNSE with the cubic nonlinearity were studied too [13]. LI was also demonstrated to control the period of Rabi oscillations and efficiency of the resonant conversion in the equation with a longitudinally modulated potential [14]. Further, the consideration of the fractional Schrödinger equation with parity-time-symmetric potentials has demonstrated that the respective symmetric bandgap structure supports diffraction-free propagation, as well as conical diffraction, in one- and two-dimensional cases [15].

As concerns FNSE, the propagation of super-Gaussian optical beams was studied in Ref. [16], using the value of LI, $\alpha$, to tune nonlinear effects [16]. In FNSE with the Kerr (cubic) nonlinearity, all solitons are known to be stable for $1 < \alpha \leq 2$, but suffer collapse (catastrophic self-compression) at $\alpha = 1$ [17]. Surface gap solitons in FNSE with defocusing Kerr nonlinearity were also studied and shown to be stable in a finite gap [18]. Recently, double-peak solitons in FNSE with a parity-time symmetric potential were found, and it was concluded that variation of LI can change their stability [19]. Two- and three-peak solitons were reported in FNSE with the Kerr nonlinearity periodically modulated in space. In the latter case, profiles and stability of the solitons are controlled by LI [20]. Three-peak and four-peak gap solitons were studied in a model with the defocusing Kerr nonlinearity and a lattice potential, their stability region shrinking with the decrease of LI [21]. Higher-order stable spatially even solitons with a truncated-Bloch-wave shape (four-peak, six-peak and eight-peak ones) were also found in FNSE with a lattice potential [22]. In FNSE with fractional-space lattices [23], vortex solitons with topological charge 1 were predicted. They are stable in an intermediate region of values of the propagation constant, which alters with variation of LI [24].

The above-mentioned results were obtained in FNSE with periodic lattice potentials. The HO potential was also found to support trapped ground states (GSs) and vortex modes [25,26]. The existence region for the GSs and vortices with topological charge 1 was identified in the model with the self-attractive nonlinearity [27]. Further, an attractive defect (represented by a delta-functional potential) inserted in the one-dimensional (1D) medium with the critical (quintic) or supercritical self-focusing nonlinearity, stabilizes a family of solitons against the collapse [28] (the fractional linear Schrödinger equation with the delta-functional potential was considered too [29]). Spontaneous symmetry breaking of stationary states in the cubic FNSE with a symmetric double-well potential was recently analyzed in Ref. [30].

In this work, we address existence and stability of single- and multi-peak confined states in FNSE with the HO potential, at values of LI corresponding to the critical ($\alpha = 1$) and supercritical ($\alpha < 1$) collapse. We find that Hermite-Gaussian polynomials provide an analytical approximation for multi-peak solutions. They may be stable in a range of values of LI that correspond to the critical and supercritical collapse, somewhat similar to the above-mentioned results reported for the usual (non-fractional) 1D nonlinear Schrödinger equation with the attractive defect and critical or supercritical self-focusing term [28].

## 2. The model

We adopt the following scaled form of FNSE with the cubic-quintic nonlinearity and HO potential, written in terms of the light propagation along axis $z$ in a planar waveguide with transverse coordinate $x$ [23-25]:

$$iu_z - \frac{1}{2}\left(-\frac{\partial^2}{\partial x^2}\right)^{\frac{\alpha}{2}} u + k_3|u|^2 u + k_5|u|^4 u - \frac{1}{2}\Omega^2 x^2 u = 0, \qquad (1)$$

where $\alpha$ is LI, $\text{sign}(k_{3,5}) = +1$ or $-1$ defines, severally, the focusing or defocusing nonlinearity, and $\Omega^2$ is the strength of the trapping HO potential. The fractional derivative in Eq. (1) is realized as the integral operator produced by the direct and inverse Fourier transforms [1,8]: $\left(-\frac{\partial^2}{\partial x^2}\right)^{\alpha/2} u = \frac{1}{2\pi}\iint dp d\xi |p|^\alpha \exp[ip(x-\xi)]u(\xi)$. Stationary solutions are looked for as $u(x,z) = \exp(i\mu z)U(x)$, where $\mu$ is the propagation constant, and real function $U$ satisfies the equation

$$\mu U + \frac{1}{2}\left(-\frac{d^2}{dx^2}\right)^{\frac{\alpha}{2}} U - k_3|U|^2 U - k_5|U|^4 U + \frac{1}{2}\Omega^2 x^2 U = 0. \qquad (2)$$

Solutions are naturally characterized by the integral power $P \equiv \int_{-\infty}^{+\infty} U^2(x) dx$.

In the limit of $|x| \to \infty$, straightforward consideration of the linearized version of Eq. (2) yields an asymptotic form of localized solutions:

$$U(x) = \text{const} \cdot \exp\left(-\frac{\Omega^{2/\alpha}}{1+2/\alpha}|x|^{1+2/\alpha}\right) \tag{3}$$

In the case of $\alpha = 2$, it carries over into the commonly known GS wave function confined by the HO potential, $U(x) = \text{const} \cdot \exp(-\Omega x^2/2)$, while at all values of $\alpha < 2$ Eq. (3) produces a super-Gaussian asymptotic form. In fact, the asymptotic form (3) may be used as an approximation for the entire ground-state (GS) solution of the linearized version of Eq. (2). Its comparison with the numerically found counterpart is displayed in Fig. 1, for $\alpha = 1.5$ and $\alpha = 1$. It is seen that the analytical approximations is relevant, although its accuracy gradually deteriorates with the decrease of the LI.

Another simple exact result for the linear limit of Eq. (2) is that all its eigenvalues scale, with variation of $\Omega$, as

$$\mu \sim \Omega^{2\alpha/(2+\alpha)}. \tag{4a}$$

The same scaling remains valid for both cubic-only and quintic-only nonlinearity, i.e., with $k_5 = 0$ or $k_3 = 0$, respectively. In these cases, the related scaling of the integral power is

$$P_{\text{cubic}} \sim \Omega^{2(\alpha-1)/(2+\alpha)}, \quad P_{\text{quint}} \sim \Omega^{(\alpha-2)/(2+\alpha)}. \tag{4b}$$

The fact that the scaling exponent for $P_{\text{cubic}}$ vanishes at $\alpha = 1$ implies, as mentioned above, that this value is critical for Eqs. (1) and (2) with the cubic nonlinearity. Similarly, the vanishing of the scaling exponent for $P_{\text{quint}}$ at $\alpha = 2$ corresponds to the well-known fact that the quintic self-focusing is critical for the usual 1D nonlinear Schrödinger equation [31,32]. The exact relations given by Eqs. (4a) and (4b) completely agree with respective numerical results.

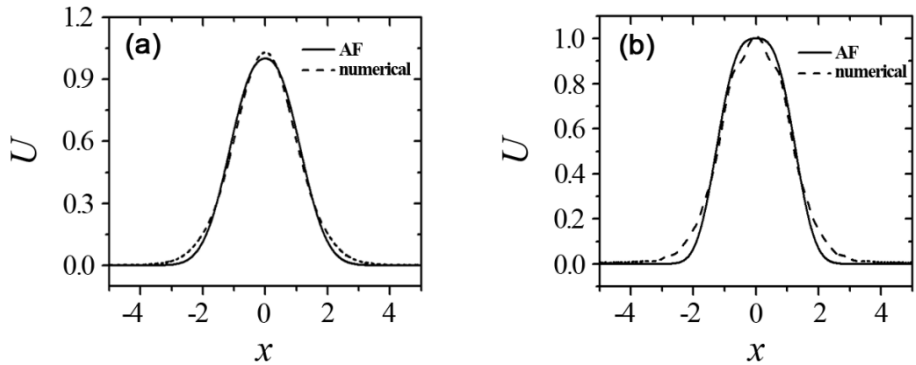

Fig.1. Comparison of the shape of the approximate solution to the linear version of Eq. (2), predicted by the asymptotic form (3) ("AF"), with a numerical solution of the same equation, at $\Omega = 1$ for $\alpha = 1.5$ (a) and $\alpha = 1.0$ (b).

In previous works, solitons were looked for as solutions to Eq. (1) with the cubic nonlinearity ($k_3 > 0, k_5 = 0$) at values of LI taken in interval $1 < \alpha \leq 2$, because the critical and supercritical collapse occur, respectively, at $\alpha = 1$ and $\alpha < 1$ [17, 31], hence at $\alpha \leq 1$ the solitons are completely unstable in the free space. Addressing the model based on Eq. (1), we aim to stabilize solitons trapped in the HO potential, in spite of the possibility of the collapse. To this end, we start by consideration of dependence of propagation constant $\mu$ on HO strength $\Omega^2$ (recall that, in the linear version of the equation, the dependence takes the exact form given by Eq. (4)). Multiplying Eq. (2) by $U$ and integrating the result in the whole spatial domain, we obtain

$$\mu \int_{-\infty}^{+\infty} U^2(x)dx + \frac{1}{4\pi} \iiint dpdxdx' |p|^\alpha \exp[ip(x-x')]U(x')U(x) +$$

$$\frac{1}{2}\Omega^2 \int_{-\infty}^{+\infty} x^2 U^2(x)dx - k_3 \int_{-\infty}^{+\infty} U^4(x)dx - k_5 \int_{-\infty}^{+\infty} U^6(x)dx = 0. \quad (5)$$

Approximate localized solutions of Eq. (2) can be looked for in the form of the ansatz suggested by eigenstates of the usual Schrödinger equation with the HO potential, i.e., as Hermite-Gauss functions with inverse squared width $a$,

$$U_n(x) = Ae^{-ax^2/2}H_n(x), n = 0,1,2,\cdots, \quad (6a)$$

where $H_n(x) = (-1)^n e^{x^2/2} \frac{d^n(e^{-x^2/2})}{dx^n}$, and the integral power of the ansatz is

$$P = \begin{cases} 2^n n! \sqrt{\pi} A^2, & a = 1 \\ 2^n A^2 a^{-2n-1}(1-a)^n \Gamma\left(\frac{2n+1}{2}\right) F\left(-n, n; \frac{1-2n}{2}; \frac{a}{2(a-1)}\right), & a \neq 1. \end{cases} \quad 6(b)$$

Here $\Gamma$ is the Gamma-function, and $F(\alpha,\beta;\gamma;z)$ is the Gauss' hypergeometric function.

Inserting ansatz (6a) in integral relation (5), an approximate analytical dependence of the propagation constant $\mu$ on $\Omega^2$ can be predicted for different orders $n$ of the Hermite-Gauss modes. Here, we produce the results for the GS ($n = 0$), first ($n = 1$) and second ($n = 2$) excited state:

$$\mu = \begin{cases} -\frac{\Omega^2}{4a} - \frac{1}{2\sqrt{\pi}} \Gamma\left(\frac{\alpha+1}{2}\right) a^{\frac{\alpha}{2}} + \frac{k_3 A^2}{2\sqrt{2}} + \frac{k_5 A^4}{2\sqrt{3}}, & (n = 0) \\ -\frac{3\Omega^2}{4a} - \frac{\Gamma\left(\frac{\alpha+3}{2}\right)}{\sqrt{\pi}} a^{\frac{\alpha}{2}} + \frac{3k_3 A^2}{2\sqrt{2}a} + \frac{20k_5 A^4}{3^{\frac{5}{2}}a^2}, & (n = 1), \quad (7) \\ \frac{1}{b_1}\left(-\frac{\Omega^2 b_4}{4a} - \frac{b_2}{2\sqrt{\pi}} + \frac{k_3 A^2 b_3}{\sqrt{2}} + \frac{2^5 k_5 A^4 b_5}{\sqrt{3}}\right), & (n = 2) \end{cases}$$

where $b_1 = \frac{3}{a^2} - \frac{2}{a} + 1$, $b_3 = \frac{105}{4a^4} - \frac{30}{a^3} + \frac{18}{a^2} - \frac{8}{a} + 4$,

$$b_2 = 4\Gamma\left(\frac{\alpha+5}{2}\right)a^{\frac{\alpha-4}{2}} - 4\left(\frac{2}{a}-1\right)\Gamma\left(\frac{\alpha+3}{2}\right)a^{\frac{\alpha-2}{2}} + \left(\frac{2}{a}-1\right)^2\Gamma\left(\frac{\alpha+1}{2}\right)a^{\frac{\alpha}{2}},$$

$$b_4 = \frac{15}{a^2} - \frac{6}{a} + 1, \quad b_5 = \frac{385}{54a^6} - \frac{35}{3a^5} + \frac{175}{18a^4} - \frac{50}{9a^3} + \frac{5}{2a^2} - \frac{1}{a} + \frac{1}{2}.$$

As shown in Fig. 2(b) below, the analytical approximation produced by ansatz (6a) is quite close to its counterpart found as a numerical solution of Eq. (2) for the same $\mu$. Furthermore, the approximation predicts general results, such as the existence and stability of various trapped modes, which completely agree with properties of families of the numerical solutions.

### 3. Stability of the single-peak ground state (GS)

To examine the linear stability of localized solutions to Eq. (1), we perturb them as

$$u(x,z) = \exp(i\mu z)[U(x) + f(x)e^{\lambda z} + g^*(x)e^{\lambda^* z}], \quad (8)$$

where $\lambda$ is the instability growth rate, and $|f|, |g| \ll |U|$ are components of the eigenmode of small perturbations, the asterisk standing for the complex conjugate. The substitution of expression (8) in Eq. (1) and linearization leads to the following eigenvalue problem:

$$i\begin{pmatrix} \hat{D}_{11} & \hat{D}_{12} \\ -\hat{D}_{12}^* & -\hat{D}_{11}^* \end{pmatrix}\begin{pmatrix} f \\ g \end{pmatrix} = \lambda \begin{pmatrix} f \\ g \end{pmatrix}, \quad (9)$$

where $\hat{D}_{11} = -\frac{1}{2}\left(-\frac{\partial^2}{\partial x^2}\right)^{\frac{\alpha}{2}} - \frac{1}{2}\Omega^2 x^2 - \mu + 2k_3|U^2| + 3k_5|U^4|$ and $\hat{D}_{12} = U^2(k_3 + 2k_5|U|^2)$.

Equation (9) can be solved numerically by means the Fourier collocation method [33]. The stationary solutions are unstable if there exist eigenvalues $\lambda$ with positive real parts, otherwise the solutions are stable.

Proceeding to direct simulations of perturbed evolution of the GS in the framework of Eq. (1), generic results were produced by taking the input as given by the numerically exact stationary solution, to which random perturbations at the 5% amplitude level were added. First, we discuss the stability of the GS for $a = 1$, in the framework of this approach.

Figure 2 shows GSs trapped in the HO potential in the case of the critical collapse in the free space for the cubic-only nonlinearity ($k_5 = 0$) in Eq. (1), i.e., $\alpha = 1$, for fixed $\Omega^2 = 1.0$. The stability spectrum of a typical GS, its stationary profile, and perturbed evolution are displayed in panels (a), (b), and (c), respectively, showing that the state is stable. Results for a subfamily of stable GS solutions, in which the approximate solutions, based on Eqs. 6(a,b) and (7), are closest to their numerically found counterparts [see details in the caption to Fig. 2(d)], are summarized in panel (d) by means of the $P(\alpha)$ dependence, for both $k_5 = 0$ and $k_5 < 0$. The plot for nonzero $k_5$ is included to demonstrate robustness of the results against the addition of the quintic nonlinearity. Note that, for given $\alpha$, GS solutions may exist with different values of total power $P$ [in other words, with different values of amplitude $A$ and squared inverse width $a$ in the corresponding ansatz (6a)]. The subfamily represented by panel (d) is selected by demanding that the amplitude of the numerically found GS solution, for given $P$, is equal to the amplitude of the approximate solution, predicted by Eqs. (6b) and (7) for the same $P$. In other words, this is a subfamily for which the analytical approximation is closest to its numerical counterpart.

It is thus found that, in the framework of Eq. (1) with the cubic-only self-focusing nonlinearity ($k_3 > 0, k_5 = 0$), the single-peak GSs are stable in the interval of LI values

$$0.7 < \alpha \leq 2, \qquad (10)$$

see Fig. 2(d). We stress that it includes region $0.7 < \alpha \leq 1$, in which the collapse takes place, making all the solitons completely unstable in the free space [17]. The stability of the GS in this region is provided by the trapping HO potential.

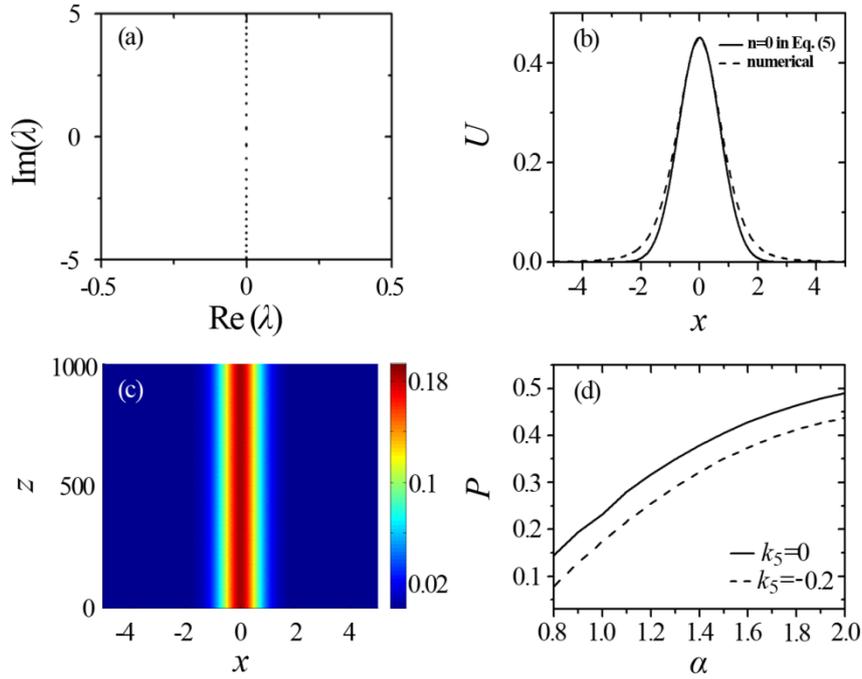

Fig. 2 (color online). Panels (a-c) display a generic fundamental single-peak mode (GS) trapped in the HO potential, in the case of the critical collapse, $\alpha = 1$, for fixed $\Omega^2 = 1.0, k_3 = 0.5$ and $k_5 = 0$ (the cubic-only nonlinearity). (a) The stability spectrum; (b) the profile of the approximate solution given by analytical ansatz (6a) (the solid line) and its numerically found counterpart (the dashed line) for $\mu = -0.42$; (c) the perturbed evolution of the stable GS for $\mu = -0.42$, initiated with the numerically exact stationary solution. (d) The total power of families of stable GSs vs. the LI, $\alpha$, for values of the quintic-nonlinearity coefficient $k_5 = -0.2$ and $0$ (the dashed and solid lines, respectively), with $k_3 = 0.5$ and $\Omega^2 = 1.0$.

## 4. Two-peak states (dipole modes)

Figure 3 illustrates properties of dipole states, produced by Eq. (2) with $k_3 > 0, k_5 = 0$ and input given by ansatz (6a) with $n = 1$ and $a = 1$. These modes, which feature a double peak in terms of the local intensity, $|U(x)|^2$, represents the first excited state created on top of the GS corresponding to $n=0$. An example of a stable dipole is displayed in Figs. 3(a-c). In the region of $\alpha < 1$, where, as said above, all self-trapped modes are destabilized by the collapse in the absence of the trapping potential, the two-peak states are found to be stable in the interval

$$0.8 < \alpha \leq 2 \tag{11}$$

for fixed $\Omega^2 = 0.5$, cf. stability interval (10) for the GS with $\Omega^2 = 1$. It includes a region of $0.8 < \alpha \leq 1$, in which all localized states are unstable in the free space, as said above. An example of an unstable dipole mode is shown in Figs. 3(d-f).

The dependence of the integral power $P$ on $\alpha$, for the stable subfamily of dipole states, defined in the same way as it was done above for GS solutions [see Fig. 2(b) and the caption to it], is displayed in Fig. 4. It is seen that $P$ first increases with the increase of LI $\alpha$, attaining a maximum value, close to 1, and then gradually decreases.

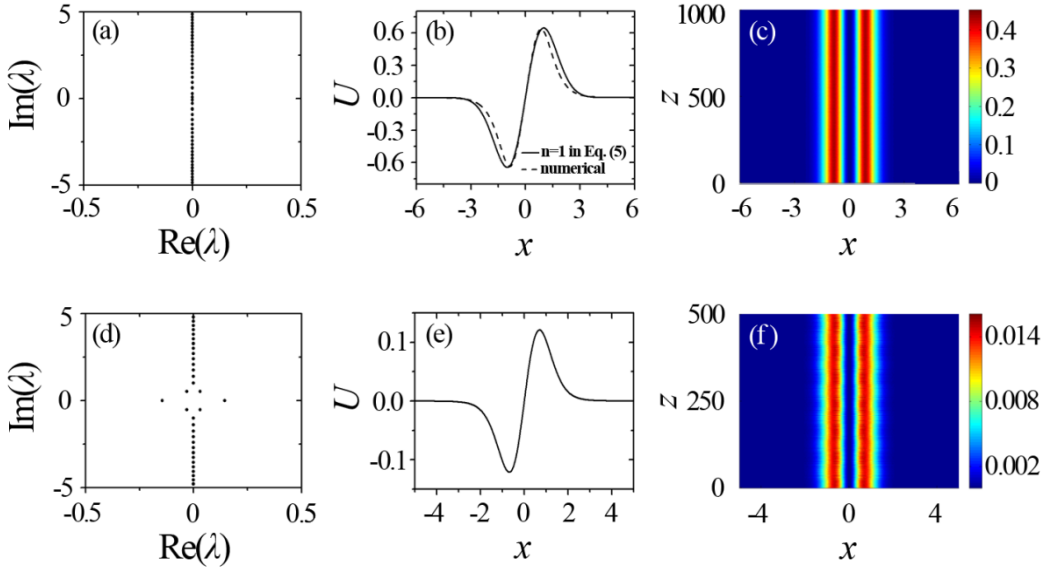

Fig. 3 (color online). Dipole (two-peak) states trapped in the HO potential, initiated by ansatz (6a) with $n = 1$, in the case of the critical collapse, $\alpha = 1$, for $\Omega^2 = 0.5$. (a)

The linear-stability eigenvalue spectrum; (b) the profile of the solution produced by the analytical approximation (the solid line) and its numerically found counterpart (the dashed line) for $\mu = -0.79$; (c) the perturbed evolution of the stable dipole mode for $\mu = -0.79$. (d-f) The eigenvalue spectrum, profile, and perturbed evolution of an unstable dipole state, for $\alpha = 1, \Omega^2 = 1.0$ and $\mu = -1.16$. In both cases shown in this figure, $k_3 = 0.5$ and $k_5 = 0$ (the cubic-only self-focusing nonlinearity).

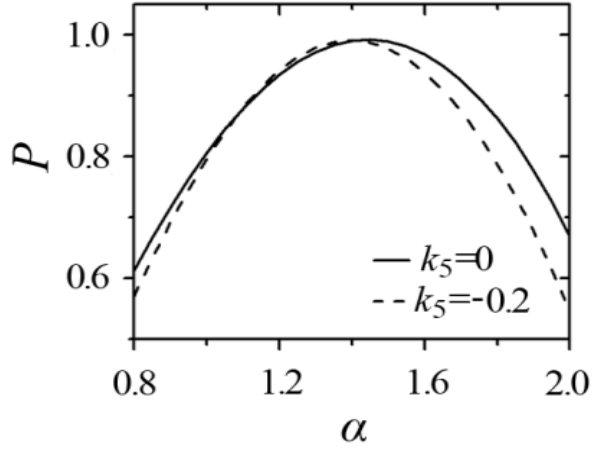

Fig. 4. The integral power of the stable dipole states vs. LI, $\alpha$, for fixed $\Omega^2 = 0.5$ and $k_3 = 0.5$ with $k_5 = -0.2$ (the dashed line) and $k_5 = 0$ (the solid line). The subfamily of the dipole solutions represented by this curve is selected in the same way as explained above for GS states, in the connection to Fig. 2(d).

## 5. Higher-order states (multipoles)

In addition to the GS and dipole solutions considered above, Eq. (1) supports a vast variety of higher-order states corresponding to $n \geq 2$ in ansatz (6a) which feature $n$ zero-crossing points, and $n+1$ local maxima of $|U(x)|^2$. An example of a stable "tripole" (three-peak mode) is shown, for $\alpha = 1.4$, in Figs. 5(a-c).

The dependence $P(\alpha)$ for the stable subfamily of tripole states, defined in the same way as it was done above for GS and dipole solutions [see Figs. 2(b) and 4] is displayed in Fig. 5(d). The $P(\alpha)$ curve attains a maximum value $P_{\max} = 1$ at $\alpha = 1.3$, and then gradually decreases, similar to the non-monotonous curve for the dipole states in Fig. 4.

By adjusting the value of LI, we can obtain modes with different intensity distributions. The stability region of the tripoles ($n = 2$) is

$$0.85 < \alpha \leq 2 \quad (12)$$

for $k_3 = 1.0$, $k_5 = -1.0$, cf. Eqs. (10) and (11). It includes interval $0.85 < \alpha \leq 1$ in which, as said above, all the self-trapped states are unstable in the free space. At other values of strength $\Omega^2$ of the HO potential, the tripole state may be unstable, see an example in Figs. 5(g-i).

Different from the GS and dipole states, the higher-order modes with $n \geq 2$ cannot be stable *unless the quintic self-defocusing is present*, to compete with the cubic focusing term. This conclusion is qualitatively similar to that reported in Ref. [27], where stabilization of trapped GSs and vortex modes was considered in the framework of the 2D nonlinear Schrödinger equation including the cubic self-attraction term and isotropic HO potential. Namely, in that case stable are the full family of GS solutions, and a part of the family of vortex modes with topological charge $S = 1$ (counterparts of the dipole modes in our model), while all higher-order vortices with $S \geq 2$, that correspond to multipoles in the present case, remain completely unstable, but the quintic defocusing term readily stabilizes the vortex states with $S \geq 2$ even in the free 2D space [34].

For fixed values of the cubic and quintic coefficients, $k_3 = -k_5 = 1.0$, and a fixed chemical potential, $\mu = -1.195$, a subfamily of stable tripoles, defined in the same way as in Fig. 5(d) [i.e., similar to how it was adopted above for the subfamilies of GS and dipole modes, in Figs. 2(d) and 4, respectively], but for the strength of the HO potential, $\Omega^2$, considered as a function of LI, $\alpha$, is presented in Fig. 6. It is seen that this dependence is a monotonously decaying one.

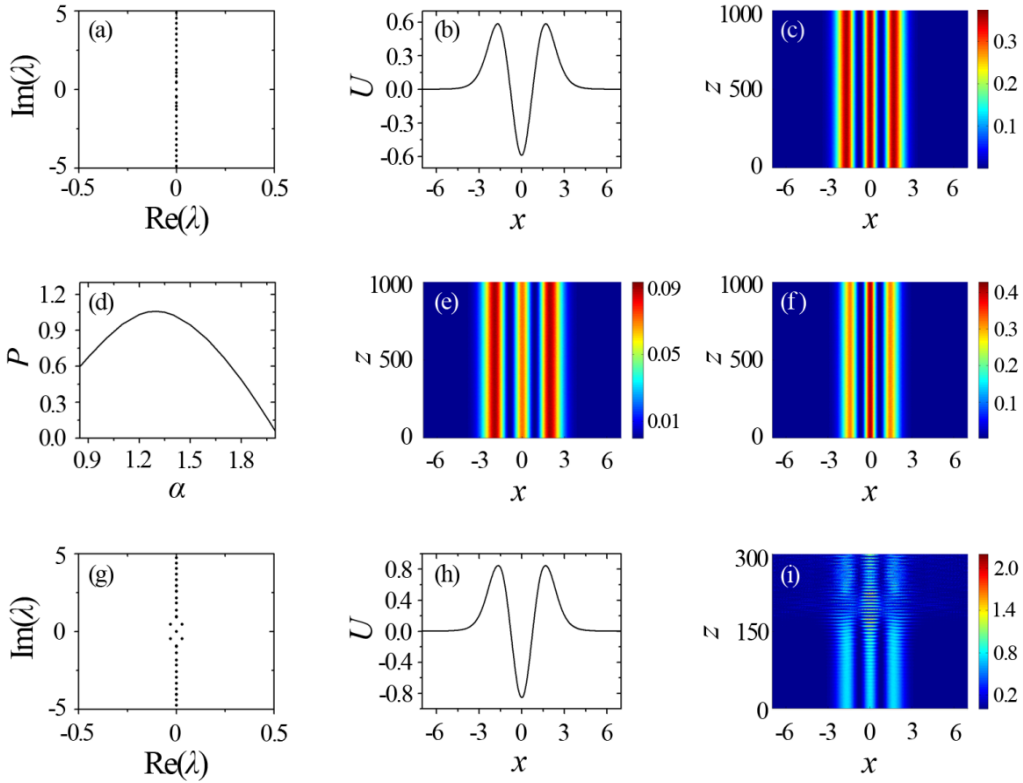

Fig. 5 (color online). Stable and unstable tripoles (three-peak modes). (a), (b) and (c): the linear-stability spectrum, profile, and stable perturbed propagation of the mode for

$\alpha = 1.4$, $\Omega^2 = 0.4$, and $\mu = -1.195$. (d) The integral power of the stable subfamily of the tripoles vs. LI, $\alpha$, for $\Omega^2 = 0.4$, defined in the same way as for the GS and dipole states in Figs. 2(d) and (4), respectively. Panels (e) and (f) display stable evolution of the tripole for $\alpha = 1.9$, $\Omega^2 = 0.4$, $\mu = -1.495$, and $\alpha = 1.0$, $\Omega^2 = 0.4$, $\mu = -1.025$, respectively. (g-i): The stability-eigenvalue spectrum, profile and perturbed evolution of an unstable tripole for $\alpha = 1.4$, $\Omega^2 = 0.43$, and $\mu = -1.2$. In all the cases, other parameters are $k_3 = 1.0$ and $k_5 = -1.0$.

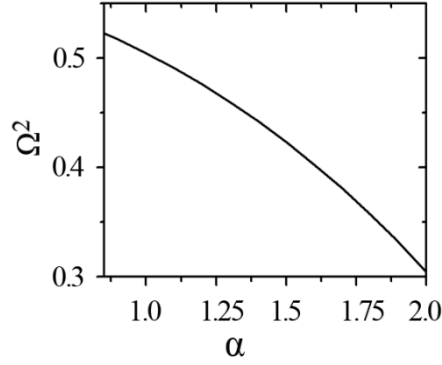

Fig. 6. (color online) The subfamily of stable tripoles, defined similar to how it is adopted in Fig. 5(d), but displayed by means of the dependence of $\Omega^2$ vs. $\alpha$, for $\mu = -1.2, k_3 = 1.0$ and $k_5 = -1.0$. Recall that the tripoles cannot be stable unless the quintic defocusing term, represented by $k_5 < 0$, is present in Eq. (1).

Higher-order states with the number of peaks $n+1$ from four to seven, for $a = 0.8$, are displayed in Fig. 7. In particular, panels(a)-(c) represent a stable four-peak mode for $\alpha = 1.4$, $\Omega^2 = 1.03, \mu = -2.5$. Further, Figs. 7(d)-(f) show stable perturbed evolution of five-peak ($\alpha = 1.2, \Omega^2 = 1.0, \mu = -2.8$), six-peak ($\alpha = 0.9, \Omega^2 = 1.0, \mu = -2.9$) and seven-peak($\alpha = 0.91, \Omega^2 = 1.03, \mu = -2.8$) modes, respectively. Note, in particular, that the latter two cases represent a multi-peak mode which is stable at $\alpha < 1$, which is impossible in the absence of the trapping potential. In terms of the local intensity, $U^2$, all these modes are symmetric about $x = 0$, but different lobes ("poles") have different amplitudes and widths. In particular, the central lobes get narrower with the decrease of LI and increase of the number of peaks.

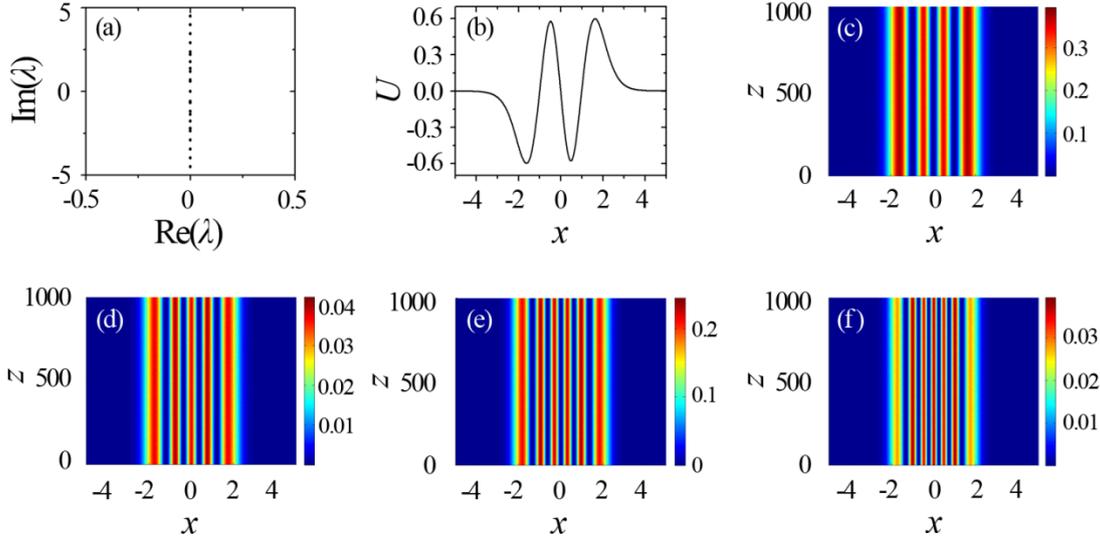

Fig. 7. (color online) Stable multi-peak states produced by Eq. (2) with $k_3 = 1.0$ and $k_5 = -1.0$. (a)-(c) The profile, linear-stability spectrum, and the stable perturbed propagation of the four-peak mode at $\alpha = 1.4, \Omega^2 = 1.03, \mu = -2.5$. (d)-(f) The stable evolution of five-peak ($\alpha = 1.2, \Omega^2 = 1.0, \mu = -2.8$), six-peak ($\alpha = 0.9, \Omega^2 = 1.0, \mu = -2.9$) and seven-peak ($\alpha = 0.91, \Omega^2 = 1.03, \mu = -2.8$) modes, respectively.

## 5. CONCLUSION

We have investigated the existence and stability of the GS (ground state), dipole mode (the first excited state), and higher-order ones in the framework of FNSE (fractal nonlinear Schrödinger equation), which is characterized by its LI (Lévy index), $\alpha$, and includes the HO (harmonic-oscillator) trapping potential. The nonlinearity is represented by cubic self-focusing, and it may also include the quintic defocusing term. The GS and dipole modes are stable with or without the quintic term. An important finding is that their stability regions extend, respectively, up to $\alpha = 0.7$ and $\alpha = 0.8$, while all self-trapped modes (solitons) in the free space (in the absence of the trapping potential) are destabilized by the critical or supercritical collapse at $\alpha = 1$ and $\alpha < 1$, respectively. If the quintic self-defocusing term is included, higher-order modes, with the number of local-intensity peaks from three up to seven, also have their stability regions in the model, including, in particular, an interval of values of LI $0.85 < \alpha < 1$ for the three-peak states [see Eq. (12)], i.e., they all may be stabilized against the supercritical collapse.

As an extension of the present work, it may be relevant to consider the stabilization of localized modes under the action of the confining potential of other types, such as a delta-functional one, cf. Ref. [28].

**Acknowledgments**

This work was supported by the National Natural Science Foundations of China (Grant No. 61675001), National Theoretical Physics Program (Grant No.11947103),the Guangdong Province Nature Foundation of China (Grant No. 2017A030311025), Characteristic Innovation Projects of General Colleges and Universities in Guangdong(Grant No.2019 KTSCX083). The work of B.A.M. is supported, in part, by the Israel Science Foundation through grant No. 1286/17.